\documentclass[11pt,a4paper]{article}
\usepackage[dvips]{graphicx}
\title{
Vibrational energy relaxation in proteins
} 
\author{Hiroshi FUJISAKI\footnote{fujisaki@bu.edu} 
and John E. STRAUB\footnote{straub@bu.edu 
}
\\
\\
Department of Chemistry, Boston University, 590 Commonwealth Ave.,
\\
Boston, Massachusetts, 02215, USA
}
\begin{document}
\maketitle

\begin{abstract}

An overview of theories 
related to vibrational energy relaxation (VER)
in proteins is presented.
VER of 
a selected mode in cytochrome c 
is studied using two theoretical approaches. One is the equilibrium simulation approach 
with quantum correction factors, and the other is the reduced model approach which 
describes the protein as an ensemble of normal modes interacting through nonlinear coupling elements.
Both methods result in estimates of the VER time (sub ps) for a CD stretching 
mode in the protein at room temperature.  The theoretical predictions are in accord with the experimental data of Romesberg's group.   A perspective on future directions for the detailed study of time scales and mechanisms for VER in proteins is presented.

Classification: Physical Science, Biophysics.

Abbreviations: VER, vibrational energy relaxation; LTZ, Landau-Teller-Zwanzig;
Mb, myoglobin; cyt c, cytochrome c; QCF, quantum correction factor.

\end{abstract}


\section{Introduction}

When a protein is excited by ligand binding, ATP attachment, or laser excitation there occurs 
vibrational energy relaxation (VER).  Energy initially ``injected'' into a localized region 
flows to the rest of the protein and surrounding solvent.  VER in large molecules (including proteins) 
itself is an important problem for chemical physics \cite{BBS88,GW04}.  Even more significant is the challenge to relate VER to fundamental reaction processes,
such as a conformational change or electron transfer of a protein, associated with protein  function.  The development of a accurate understanding of VER in proteins is an essential step toward the goal of controlling protein dynamics \cite{DLZ02}.

Due to the advance of laser technology,
there have been many experimental studies of 
VER in proteins \cite{LLKH94,LJA96,MK97,SSJLA99,RKYSDC00,WYDRSCSC00,XMHA00,Fayer01,CJR01,
CJR02,YDC02,XMA02,MNTALF03,SZDR04}.
These experimental works are impressive but it is 
difficult to derive detailed information from the experimental data alone.
Theoretical approaches including atomic-scale simulations
can provide more detailed information.  In turn, experimental 
data can be used to refine simulation methods and empirical force fields.
This combination of experimental and theoretical studies of for protein structures and dynamics 
has begun to blossom.  As experimental methods develop further and theoretical approaches grow in accuracy, the relationship will become fruitful.
 
There have been many theoretical tools (Sec.~\ref{sec:theories}) 
developed to analyze VER in proteins.   Some aspects of VER in proteins 
can be explained by perturbative formulas based on the equilibrium 
condition of the bath (Sec.~\ref{sec:cytc}),  but the use of the perturbative formulas may be too restrictive to generally describe 
protein dynamics at room temperature. 
In this paper, 
we not only discuss the success of such established methods
but also present a perspective on the future study of VER in proteins.

\section{Theories}
\label{sec:theories}

In this section, we present a selective overview of theories appropriate for the study of VER in proteins.
For the most part, these theories have been developed to deal with VER in 
liquids, solids, or glasses.  The reader is referred to a number of recent reviews \cite{Leitner05,FBS05,FBS05b}. 
We refer to two distinct categories: 
one based on equilibrium dynamics and Fermi's golden rule, while the other is based on nonequilibrium dynamical models.  

\subsection{Fermi's golden rule}

If (a) there is a clear separation between the system and bath,
(b) the coupling between them is weak enough, and (c) the bath is assumed 
to be at thermal equilibrium, we can use quantum mechanical 
perturbation theory to derive a vibrational population relaxation rate through 
Fermi's golden rule \cite{FBS05,FBS05b}
\begin{eqnarray}
\frac{1}{T_1}
=
 \frac{\tanh({\beta \hbar \omega_S/2})}{\beta \hbar \omega_S/2}  
\int_0^{\infty} dt \, \cos (\omega_S t)  \, \zeta_{\rm qm}(t)
\label{eq:Fermi}
\end{eqnarray}
where the force-force correlation function $\zeta_{\rm qm}(t)$ is defined as 
\begin{equation}
\zeta_{\rm qm}(t)=
\frac{\beta}{2 m_S} \langle {\cal F}(t) {\cal F}(0) + {\cal F}(0) {\cal F}(t) \rangle_{\rm qm},
\end{equation}
${\cal F}(t)$ is the quantum mechanical force applied to the relaxing bond (system) considered,
$m_S$ is the system mass, 
$\omega_S$ is the system frequency,
$\beta$ is an inverse temperature, 
and the above bracket indicates a quantum mechanical average.

However, this time correlation function is very hard to numerically calculate.
As a result, many approximate 
schemes have been proposed to address this limitation.  
A number of the most successful approaches 
is mentioned below.  

\subsubsection{Landau-Teller-Zwanzig formula}
\label{sec:LTZ}

The most simple approximation is to take the classical limit ($\hbar \rightarrow 0$) 
of Eq.~(\ref{eq:Fermi})
\begin{equation}
\frac{1}{T_1^{\rm cl}}
=
\frac{\beta}{m_S} \int_0^{\infty} dt \cos (\omega_S t) \langle {\cal F}(t) {\cal F}(0) \rangle_{\rm cl}.
\end{equation}
Here the bracket denotes a classical ensemble average. 
This is called the Landau-Teller-Zwanzig (LTZ) formula, 
which has been applied to the study of VER in liquids \cite{WWH92}.
This strategy was used by Sagnella and Straub to discuss 
the VER of CO in Mb$^*$CO \cite{SS99}.
This approximation should be good for low frequency modes,
but it becomes questionable for high frequency modes due to quantum effects. 
As such, advanced methods have been proposed to address this 
deficiency of the LTZ formula.

\subsubsection{Quantum correction factor}
\label{sec:QCF}

The first alternative to the LTZ formula is the 
quantum correction factor (QCF) method.
The basic idea of the QCF method is 
to relate a quantum mechanical correlation function with its classical analog \cite{SP01}. 
When this is done for the force autocorrelation function in Eq.~(\ref{eq:Fermi}), 
the final expression for the VER rate $1/T^{\rm QCF}_1$ is
\begin{eqnarray}
\frac{1}{T_1^{\rm QCF}}
\simeq 
\frac{Q(\omega_S)}{Q_H(\omega_S)} 
\frac{1}{T_1^{\rm cl}} 
\label{eq:qcf}
\end{eqnarray}
where $Q(\omega_S)$ is the QCF for the VER process considered and 
$Q_H(\omega_S)$ is the QCF for a one phonon relaxation process (harmonic QCF)
\begin{equation}
Q_H(\omega)
=
\frac{\beta \hbar \omega}{1-e^{-\beta \hbar \omega}}.
\end{equation}
In the previous work \cite{FBS05,FBS05b}, this result was expressed as
$T_1^{\rm QCF}
\simeq 
[{\beta \hbar \omega_S}/ {Q(\omega_S)}]
{T_1^{\rm cl} }
$
which is correct in the limit $\beta \hbar \omega_S \gg 1$, as was appropriate for those studies.

If the relaxation process is the linear resonance (1:1 Fermi resonance), 
then $Q(\omega_S)=Q_H(\omega_S)$, i.e., $T_1^{\rm QCF}=T_1^{\rm cl}$ \cite{BB94}.
Skinner and coworkers have provided a theoretical framework for organizing and 
expanding on a variety of QCFs appropriate for specific dynamical processes, 
dependent upon the underlying mechanism of VER.
Though this strategy has been criticized \cite{SO99,MSO01,SG03a},
it is known that 
the QCF method works rather well for specific problems \cite{LNMS04,FBS05,FBS05b}.

\subsubsection{Reduced model approach}
\label{sec:reduced}

An alternative approach to address the shortcomings of the LTZ formula is 
to use the reduced model approach \cite{Leitner05,FBS05,FBS05b}, which exploits a normal mode picture of the protein.  By representing the Hamiltonian in terms of system, bath, and interaction terms
\begin{eqnarray}
{\cal H} &=& 
{\cal H}_S+{\cal H}_B+{\cal V}_3 +{\cal V}_4 + \cdots,
\\
{\cal H}_S 
&=& \frac{p_S^2}{2}+\frac{\omega_S^2}{2}q_S^2,
\\
{\cal H}_B 
&=& \sum_k \frac{p_k^2}{2}+\frac{\omega_k^2}{2}q_k^2,
\label{eq:4ath}
\end{eqnarray}
the residual interaction term may be expanded perturbatively as 
\begin{eqnarray}
{\cal V}_3 
&=&
\frac{1}{3} \sum_{k,l,m} G_{klm} q_k q_l q_m,
\\
{\cal V}_4 
&=&
\frac{1}{4} \sum_{k,l,m,n} H_{klmn} q_k q_l q_m q_n.
\label{eq:4bth}
\end{eqnarray}
Calculating the force from this Hamiltonian,
and substituting it into Fermi's golden rule Eq.~(\ref{eq:Fermi}),
we can derive a lowest order VER rate as \cite{FBS05,FBS05b}
\begin{eqnarray}
\frac{1}{T_1}
\simeq 
\frac{\tanh (\beta \hbar \omega_S/2)}{\hbar \omega_S}
\sum_{k,l} 
\left[
\frac{\gamma \zeta^{(+)}_{k,l}}{\gamma^2+(\omega_k+\omega_{l}-\omega_S)^2}
+
\frac{\gamma \zeta^{(+)}_{k,l}}{\gamma^2+(\omega_k+\omega_{l}+\omega_S)^2}
\nonumber
\right.
\\
\left.
+
\frac{\gamma \zeta^{(-)}_{k,l}}{\gamma^2+(\omega_k-\omega_{l}-\omega_S)^2}
+
\frac{\gamma \zeta^{(-)}_{k,l}}{\gamma^2+(\omega_k-\omega_{l}+\omega_S)^2}
\right]
\label{eq:rate}
\end{eqnarray}
where 
\begin{eqnarray}
\zeta_{k,l}^{(+)}
&=&
\frac{\hbar^2}{2}
\frac{(G_{S,k,l})^2}{\omega_k \omega_{l}}
(1+n_k +n_{l} + 2 n_k n_{l}),
\\
\zeta_{k,l}^{(-)}
&=&
\frac{\hbar^2}{2}
\frac{(G_{S,k,l})^2}{\omega_k \omega_{l}}
(n_k +n_{l} + 2 n_k n_{l}),
\\
n_k &=& 1/(e^{\beta \hbar \omega_k}-1). 
\end{eqnarray}
and in previous papers \cite{FBS05,FBS05b}, $m_S$ in the perturbative formulas should read $m_S=1$ as mass-weighted coordinates were employed.

The original formula contains delta functions, 
and we have included a width parameter $\gamma$ to broaden the delta functions 
for numerical calculations.
There exists another well known formula to describe 
the VER rate, the Maradudin-Fein formula \cite{MF62,Leitner05},
\begin{eqnarray}
W &=& W_{\rm decay} + W_{\rm coll},
\label{eq:MF}
\\
W_{\rm decay} 
&=&
\frac{\hbar}{2 \omega_S} 
\sum_{k,l} \frac{(G_{S,k,l})^2}{\omega_k \omega_{l}}
(1+n_k +n_{l}) \frac{\gamma}{\gamma^2+(\omega_S -\omega_k-\omega_{l})^2},
\\
W_{\rm coll}
&=&
\frac{\hbar}{\omega_S} 
\sum_{k,l} \frac{(G_{S,k,l})^2}{\omega_k \omega_{l}}
(n_k -n_{l}) \frac{\gamma}{\gamma^2+(\omega_S +\omega_k-\omega_{l})^2}
\end{eqnarray}
with a width parameter $\gamma$.
These two formulas are numerically similar for small $\gamma$, and 
equivalent for the limit of $\gamma \rightarrow 0$ \cite{KTF94}.
This is a quantum mechanically exact  treatment given the approximate truncated form of the interaction Hamiltonian.  We have found that the truncation error 
(the contribution from higher order terms) 
can be a serious problem, especially for proteins.
For a more accurate treatment of VER, we must appeal to more advanced methods, described below.

\subsubsection{Other (advanced) approaches}

Methods that complement the above three methods 
involve calculating the force auto correlation function
$\zeta(t)$ appearing
in Fermi's golden rule 
using different levels of approximations.
Shi and Geva \cite{SG03a} used a semiclassical approximation \cite{Miller01} for 
$\zeta(t)$, and showed that even the slow relaxation of 
neat liquid oxygen (at 77K) can be well reproduced by their method.
From their study, it was shown that the short time dynamics of 
$\zeta(t)$ is important to predict the correct VER rate. 
 This implies that the short time approximation may be adequate for an accurate estimate of $\zeta(t)$. 
Various time-dependent self-consistent field methods \cite{JG99} 
or path integral methods \cite{Makri99} should be applicable to calculate $\zeta(t)$. 
For other methods, the reader is referred to additional works \cite{RR01,KR02,PNR03}.

To derive Fermi's golden rule, 
we have used the Bader-Berne correction \cite{BB94}, 
which holds only for harmonic systems.
Bader, Berne, Pollak, and H\"anggi extended 
this to an anharmonic system within a classical framework \cite{BBPH96},
and found that the VER of such a system can be nonexponential in time 
and is significantly affected by the character of the bath.
This consideration will be important when one studies the VER of CO in Mb,
especially for the VER of a highly excited CO bond.
 
\subsection{Nonequilibrium simulation}

The above equilibrium simulation methods based on Fermi's golden rule 
invoke several assumptions as described above.
These assumptions might be invalid in some cases.  As VER is a nonequilibrium phenomenon, the appeal of nonequilibrium approaches is quite natural.

\subsubsection{Classical approaches}

Classical nonequilibrium simulations to 
investigate VER in proteins were first conducted by Henry, Eaton and Hochstrasser \cite{HEH86}.
In conjunction with their experimental studies, 
they employed classical molecular dynamics simulations of heme cooling in Mb and cyt c in vacuum and found that heme cooling occured on two time scales: short (1-4 ps) and long (20 ps for Mb and 40 ps for cyt c).
Nagaoka and coworkers carried out the similar simulations for Mb in vacuum and 
obtained similar time scales \cite{OHN01}. Importantly, they found
that the normal mode frequencies localized in the proprionate side chains of the heme 
are resonant with the water vibrational frequencies. 

Straub's group executed several numerical simulations for Mb and cyt c in water.
Sagnella and Straub showed that the VER for Mb in water can be described 
by a single exponential with a few ps VER time \cite{SS01}. 
Furthermore, they suggested that the main doorway of VER is 
due to the coupling between the proprionate side chains and water,
which is in accord with Nagaoka's and Hochstrasser's observations.
Bu and Straub supported this view through simulations of mutant Mb's and Mb variants having structurally modified heme groups \cite{BS03a}.
They also investigated VER of cyt c in water, and found 
that the VER presents a biphasic exponential decay with 
two VER times: fast (a few ps) and slow (tens of ps) \cite{BS03b}.

Kidera's group studied VER in proteins from a different perspective \cite{MMK03}.
They excited a single normal mode in Mb, and examined the vibrational 
energy transfer (VET) between normal modes.
As is well known, VET is caused by (nonlinear) 
Fermi resonance: if the frequency matching is good, and the 
coupling between normal modes is strong enough, there will be VET. 
This picture is very useful to characterize VET at low temperatures.  However,
at high temperature there occurs non-resonant VET.   They numerically found that the amount of VET is
proportional to a reduced model energy including up to third 
order coupling elements (see also Sec.~\ref{sec:reduced}).


\subsubsection{Quantum approaches}
\label{sec:quantum}

For all but the simplest systems, quantum approaches for nonequilibrium simulations are 
approximate and time-consuming.   Nevertheless,
these methods can overcome problems in inherent to classical simulations.  
There are two categories: vibrationally quantum methods, 
and electronically quantum ones. 

Hahn and Stock used a reduced model (consisting of the 
retinal rotation and other environmental degrees of freedom)
to describe the pump-probe spectroscopy for the retinal chromophore in rhodopsin \cite{HS00}.
Flores and Batista, employing the same model, 
suggested the possibility to control the retinal rotation 
by two (chirped) laser pulses \cite{FB04}.
To solve the quantum dynamics for the large system,
they employed time-dependent 
self-consistent field (TD-SCF) methods \cite{JG99}.
Notably, vibrational SCF methods have been used to calculate 
vibrational energy levels for a small protein (BPTI) \cite{RGER95}.

The combination of classical simulations 
for vibrational motions and quantum calculations 
for electronic structure, in some portion of the molecule, has been widely used for the calculations of up to moderate-sized molecules.
One cutting edge application to a large system is 
the calculation of bacteriorhodopsin's photoisomerization 
in the excited chromophore state by Hayashi, Tajkhorshid, and Schulten \cite{HTS03}.
In their treatment,
a portion of the retinal chromophore including three double bonds 
was treated as the quantum mechanical region, and the complement, including the protein and water, as the molecular mechanical region.
During the simulations, there occurs nonadiabatic transitions betweeen 
two electronic states (S$_0$ and S$_1$) which was treated semiclassically.
They numerically showed that only one bond (C$_{13}$=C$_{14}$) rotates unidirectionally due 
to the coupling with the protein, and found that several other bonds can twist in any direction 
if there is no protein.


\section{Cytochrome c}
\label{sec:cytc}

In this section, we shall focus on one protein, cytochrome c (cyt c), 
and review the recent theoretical studies about this protein.
There are several reasons to select this protein as a 
prototypical one: 
(a) Cyt c is a relatively small protein with 1745 atoms. 
Other proteins of similar scale are Mb, BPTI, and human lysozyme.
(b) The detailed X-ray structure is known for cyt c.
(c) Cyt c has a function of electron transfer.
The basic theoretical  and computational works on cyt c 
were summarized by Wolynes and coworkers \cite{WZSMW93}.
Wang, Wong, and Rabitz studied VER in cyt c using 
their hydrodynamical method \cite{WWR98}.
Garcia and Hummer found anomalous diffusion for some principal 
components of cyt c in water \cite{GH99}.
Here we describe the results of our studies on the VER of cyt c using two different 
methods (the QCF method in Sec.~\ref{sec:QCF} and the reduced model approach in Sec.~\ref{sec:reduced}) 
and compare them with the experimental results of Romesberg's group \cite{CJR01,CJR02,SZDR04}.

\begin{figure}[htbp]
\hfill
\begin{center}
\includegraphics[scale=0.5]{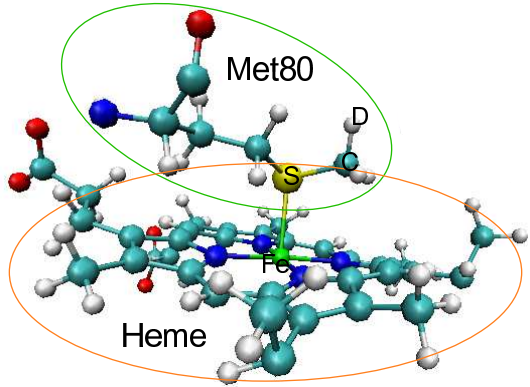}
\end{center}
\caption{
Cytochrome c near heme. Only the 80th methionine (Met80) residue and heme are depicted.
Relevant atoms (C, D, S, Fe) are also indicated. This figure was created by VMD 
(Visual Molecular Dynamics) \cite{VMD}.
}
\label{fig:cytc}
\end{figure}

\subsection{Quantum correction factor approach for cyt c}

\begin{figure}[htbp]
\hfill
\begin{center}
\begin{minipage}{.42\linewidth}
\includegraphics[scale=0.9]{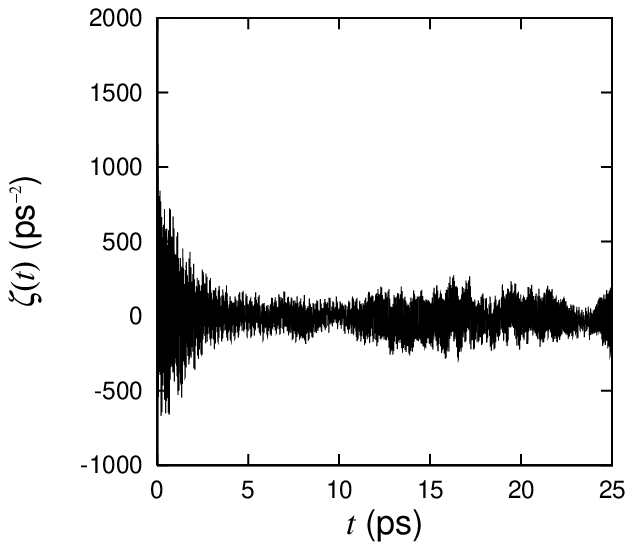}
\end{minipage}
\hspace{1cm}
\begin{minipage}{.42\linewidth}
\includegraphics[scale=0.9]{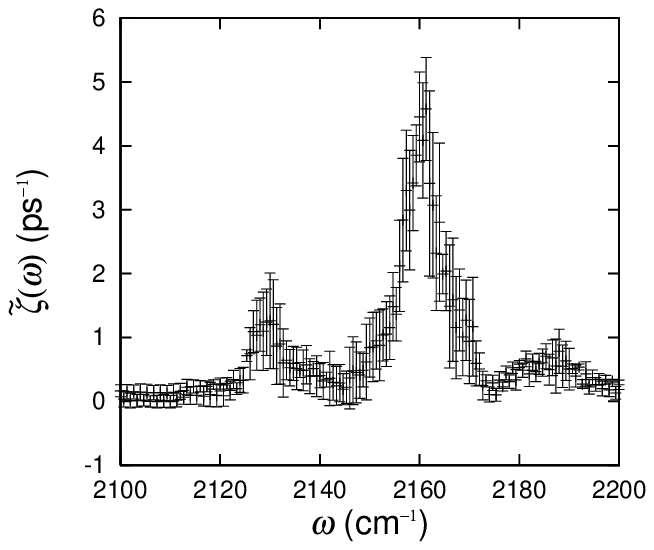}
\end{minipage}
\end{center}
\caption{
Left: Averaged force auto correlation function for four trajectories at 300K.
Right: Fourier spectrum for the four correlation functions with error bars. 
}
\label{fig:classical}
\end{figure}

Bu and Straub \cite{BS03} employed the QCF approach (Sec.~\ref{sec:QCF}) 
to estimate the VER rate of a CD bond 
in the terminal methyl group of Met80 in cyt c (see Fig.~\ref{fig:cytc}). 
Their calculations were carried out using the program CHARMM \cite{CHARMM},
and cyt c was surrounded by water molecules at 300K.
In Fig.~\ref{fig:classical}, 
we show the force autocorrelation function and its power spectrum.
With the CD bond frequency $\omega_S=2133$ cm$^{-1}$, 
we find $1/T_1^{\rm cl}=\tilde{\zeta}_{\rm cl}(\omega_S) \simeq  0.4 \sim 1.0$ ps$^{-1}$, 
so that the classical VER time is $1.0 \sim 2.5$ ps.

Since the CD bond frequency is located in a transparent 
region of the vibrational density of states, with no other state overlapping with 
this frequency \cite{BS03}, it is concluded that there is no linear resonance 
(1:1 resonance). 
To use the QCF method,
we thus need to assume nonlinear resonances 
corresponding to multiphonon VER processes. 
If the VER process assumes that two lower frequency bath modes, 
having frequencies $\omega_A$ and $\omega_S - \omega_A$, 
are each excited by one quantum of vibrational energy, the appropriate QCF 
(harmonic-harmonic QCF) is \cite{SP01}
\begin{equation}
Q_{HH}(\omega_S)
= Q_H(\omega_A) Q_H(\omega_S-\omega_A).
\label{eq:QCFHH}
\end{equation}
Alternatively, if the VER process is one 
that leads to the excitation of one bath vibrational mode of frequency $\omega_A$, 
with the remaining energy $\hbar (\omega_S - \omega_A)$ being transferred to 
lower frequency bath rotational and translational modes, the appropriate QCF 
(harmonic-harmonic-Schofield QCF) is \cite{SP01}
\begin{equation}
Q_{H-HS}(\omega_S)
=
Q_H(\omega_A) \sqrt{Q_{H}(\omega_S-\omega_A)} e^{\beta \hbar (\omega_S-\omega_A)/4}.
\label{eq:QCFHHS}
\end{equation}
We need to determine the value of $\omega_A$ to use these formulas.
From the normal mode and anharmonic coefficient calculations carried out in
Sec.~\ref{sec:reducedforcytc},
we have found that the CD mode is strongly resonant with two lower frequency modes, 
$\omega_{1655}$ = 685.48 cm$^{-1}$) and $\omega_{3823}$ = 1443.54 cm$^{-1}$), where
$|\omega_S-\omega_{1655}-\omega_{3823}| =0.03$ cm$^{-1}$ for the standard parameters
of CHARMM.
We might be able to choose $\omega_A=1443.54$ cm$^{-1}$ or 685.48 cm$^{-1}$.
If we choose $\omega_A=1443.54$ cm$^{-1}$ at 300K, 
$T_1^{\rm cl}/T_1^{\rm QCF}=Q_{HH}(\omega_S)/Q_H(\omega_S)=2.3$ for the harmonic-harmonic QCF 
and 
$T_1^{\rm cl}/T_1^{\rm QCF}=Q_{H-HS}(\omega_S)/Q_H(\omega_S)=2.8$ 
for the harmonic-harmonic-Schofield QCF.
Thus we find $T_1^{\rm QCF}=T_1^{\rm cl}/(2.3 \sim 2.8) \simeq 0.3 \sim 1.0$ ps.

\subsection{Reduced model approach for cyt c}
\label{sec:reducedforcytc}

\begin{figure}[htbp]
\hfill
\begin{center}
\begin{minipage}{.42\linewidth}
\includegraphics[scale=0.9]{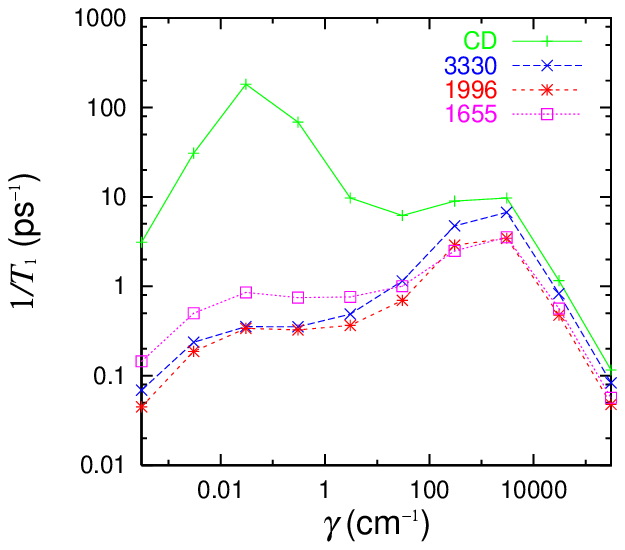}
\end{minipage}
\hspace{1cm}
\begin{minipage}{.42\linewidth}
\includegraphics[scale=0.8]{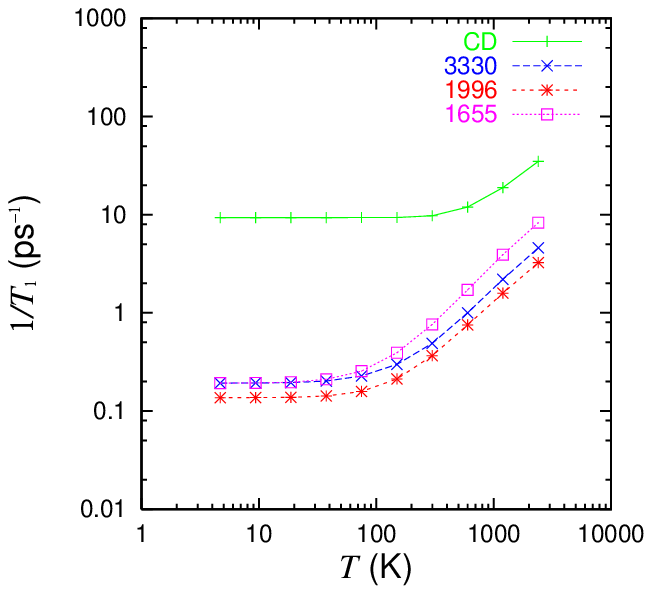}
\end{minipage}
\end{center}
\caption{
Left: VER rates 
for the CD mode 
($\omega_{CD}=2129.1$ cm$^{-1}$) 
and the other lower frequency modes 
($\omega_{3330}=1330.9$ cm$^{-1}$, $\omega_{1996}=829.9$ cm$^{-1}$, $\omega_{1655}=685.5$ cm$^{-1}$)
as a function of $\gamma$ at 300K.
Right: Temperature dependence of the VER rate for the four modes 
with $\gamma=3$ cm$^{-1}$.
}
\label{fig:VER}
\end{figure}

Fujisaki, Bu, and Straub \cite{FBS05,FBS05b} took the reduced model approach (Sec.~\ref{sec:reduced}) to 
investigate the VER for the same CD bond stretching in cyt c.
However, in their calculation, all modes represent normal modes, 
so the CD ``bond'' turned out to be the CD ``mode.''
Using the formulas in Sec.~\ref{sec:reduced}, 
they calculated the VER rate for the CD mode ($\omega_{CD}=2129.1$ cm$^{-1}$) 
and other low frequency modes 
($\omega_{3330}=1330.9$ cm$^{-1}$, $\omega_{1996}=829.9$ cm$^{-1}$, $\omega_{1655}=685.5$ cm$^{-1}$)
as a function of the width parameter $\gamma$ (Fig.~\ref{fig:VER}).
To this end, they needed to calculate anharmonic coupling coefficients
according to the formula
\begin{equation}
G_{S,k,l}=\frac{1}{2} \frac{\partial^3 V}{\partial q_S \partial q_k \partial q_l}
\simeq \frac{1}{2} \sum_{ij} U_{ik}U_{jl} 
\frac{K_{ij}(\Delta q_S)-K_{ij}(-\Delta q_S)}{2\Delta q_S}
\end{equation}
where $U_{ik}$ is an orthogonal matrix that diagonalizes 
the (mass-weighted) hessian matrix at the mechanically stable structure $K_{ij}$,
and $K_{ij}(\pm \Delta q_{S})$ is a hessian matrix 
calculated at a shifted structure along the direction of 
a selected mode with a shift $\pm \Delta q_{S}$.

If we take $\gamma \simeq \Delta \omega \sim 3$ cm$^{-1}$,
we have $T_1 \simeq 0.1$ ps, 
which agrees with the sub-picosecond time scale for relaxation predicted using the QCF method ($T_1^{\rm QCF} = 0.3 \sim 1.0$ ps).
We also see that the low frequency modes have longer VER time, a few ps,
which agrees with the similar calculations by Leitner's group \cite{Leitner05}.
In the right of Fig.~\ref{fig:VER}, we show the temperature dependence of the VER rate.
At low temperatures, the VER rate 
becomes flat as a function of temperature.
At these lower temperatures, the VER is caused by the remaining quantum fluctuation associated with zero point energy.


\subsection{Related experiment}

Here we discuss the related experiment by Romeberg's group \cite{CJR01,CJR02,SZDR04}.
They measured the shifts and widths of the spectra 
for different forms of cyt c;
the widths of the spectra (FWHM) 
were found to be $\Delta \omega_{\rm FWHM} \simeq 6.0 \sim 13.0$ cm$^{-1}$.
If we can neglect inhomogeneous effects, 
the estimate of the VER time becomes
\begin{equation}
T_1 \sim 5.3/\Delta \omega_{\rm FWHM}  \,({\rm ps})
\label{eq:estimate}
\end{equation}
which corresponds to $T_1 \simeq 0.4 \sim 0.9$ ps.
This estimate is similar to the QCF prediction 
using Eq.~(\ref{eq:qcf}) ($\simeq 0.3 \sim 1.0$ ps)
and larger than the estimate by 
the reduced model approach using Eq.~(\ref{eq:rate}) or (\ref{eq:MF}) ($\simeq$ 0.1 ps).
This might be due to the strong resonance between the three modes
(4357, 3823, 1655), which forms a peak near $\gamma \simeq 0.03$ cm$^{-1}$ 
in the left of Fig.~\ref{fig:VER}.
This resonance causes an increase in the VER rate,
so we can say that this estimate of the VER rate is too large.
On the other hand, there is no peak for the low frequency modes for $\gamma < 10$ cm$^{-1}$;
the estimate of the VER rate does not seem to be affected by the resonances.

Note also that
Romesberg's group studied
Met80-3D, methionine with three deuteriums on the terminal methyl group,
while we have examined Met80-1D, with one deuterium.
It is known that the CHARMM force field calculation does not give an accurate 
value of the absorption peak. On the other hand, the DFT calculation 
for the methionine leads to much better results 
(Matt Cremeens, private communication).
Clearly, we must improve our force field parameters according to DFT calculations,
and examine how further optimization of the parameters 
affects the resonance structures and the VER rate of the protein.

\section{Concluding Remarks}

In this paper, we have described theoretical (Sec.~\ref{sec:theories}) approaches to the study of VER in proteins. 
We have examined VER of a CD stretching bond (mode) 
in cytochrome c from the QCF approach (Sec.~\ref{sec:QCF})
and the reduced model approach (Sec.~\ref{sec:reduced}). 
For the CD mode in cyt c (in vacuum) at room temperature, 
both approaches yield similar results for the VER rate, 
which is also very similar to an estimate derived from an experiment by Romesberg's group.
Our work demonstrates both the feasibility and accuracy of a number of theoretical approaches 
to estimate VER rates of selected modes in proteins.

\begin{figure}[htbp]
\hfill
\begin{center}
\begin{minipage}{.42\linewidth}
\includegraphics[scale=0.9]{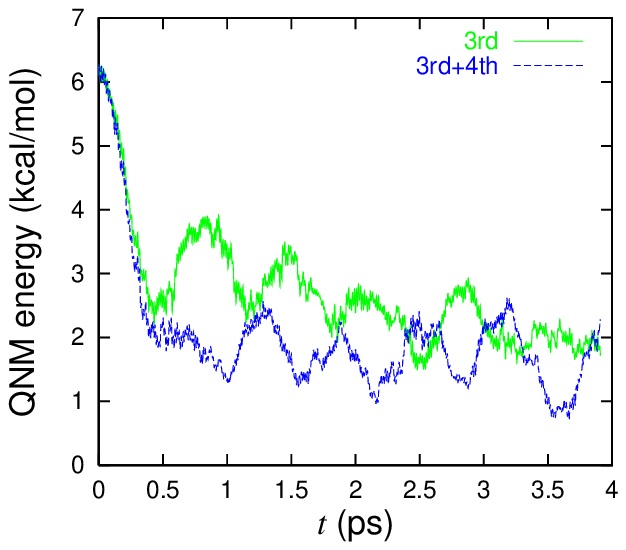}
\end{minipage}
\hspace{1.0cm}
\begin{minipage}{.42\linewidth}
\includegraphics[scale=0.9]{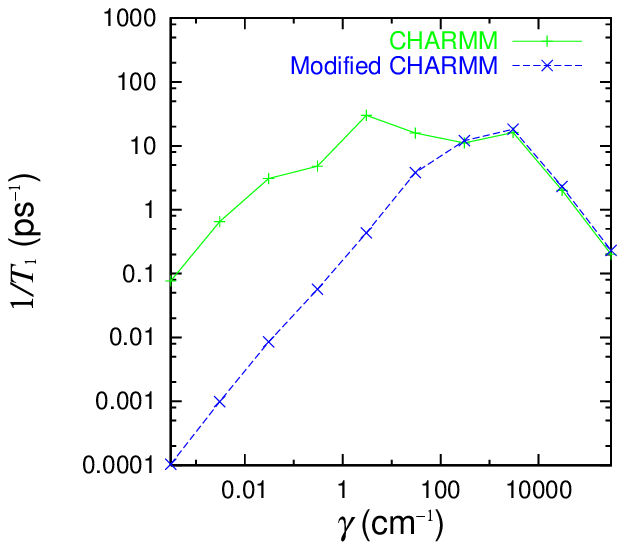}
\end{minipage}
\end{center}
\caption{
Left: Classical nonequilibrium dynamics calculation for an isolated methionine using 
the reduced model (see Sec.~\ref{sec:reduced}).
Initially the CD mode is excited to $\simeq$ 6 kcal/mol.
The other modes are excited according 
to the Boltzmann distribution at 300K. 
The quenched normal mode (QNM) energy for the CD mode (ten trajectory average) is shown for two cases: 
(a) only 3rd order coupling elements are included, 
and (b) both 3rd and 4th order coupling elements 
are included.
Right: VER rate calculation for the Met80-3D case with different force field parameters.
For the modified CHARMM parameters, 
the CD bond force constant was increased by ten percent in order to match with the absorption 
peak of the experiment by Romesberg's group.
}
\label{fig:preliminary}
\end{figure}

There are advantages and disadvantages of the (a) QCF approach and (b) reduced model approach to the prediction of VER rates in proteins.
The QCF method is simple and applicable even 
for a large molecule like a protein.
However, the VER mechanism may not be known {\it a priori}, and 
it must be supplemented by other methods such as anharmonic 
coefficient calculations. Furthermore, the method relies on 
the local mode picture, which is easily applicable for high frequency (localized) modes,
but not for low frequency (delocalized) modes. 
The reduced model approach is quantum mechanically exact,
and easily applicable for VER of low frequency modes. 
However, the anharmonic coefficient calculation is 
cumbersome even for the third order coupling terms in cyt c.  Moreover,  such a Talyor series expansion has not been shown to converge at low order
for general systems \cite{MSO01}.
Our preliminary calculations 
show that the classical VER dynamics using an isolated methionine 
does not seem to be affected by including the fourth-order coupling elements 
(see the left of Fig.~\ref{fig:preliminary}),
but we need to examine this issue further with quantum mechanical approaches
such as TD-SCF methods.
There is also an unsolved problem of the width parameter.
Actually, this problem is not peculiar to the reduced model approach.
The introduction of the width corresponds to coarse-graining, which 
also appears in the QCF approach when one averages the power spectrum 
of the force auto-correlation function.
The most ``ab initio'' approach to solve this problem 
is a rigorous quantum mechanical treatment of the 
tier structure of energy levels in the protein \cite{SM93}.
The other appealing way is to regard $\gamma$ as a hopping rate between 
conformational substates \cite{MKS91,CV95}, 
or a frequency correlation time, that may be derived from estimates of the 
frequency fluctuation \cite{MHS97,ME98}.

Since both the QCF and reduced model approaches
are based on Fermi's golden rule,
there is a limitation for the strength of the interaction 
between the system and bath.
There is a need to develop other methods without this deficiency.  
Promising approaches include nonequilibrium molecular dynamics methods \cite{NS03}, 
time-dependent self-consistent field methods \cite{JG99}, 
mixed quantum-classical methods \cite{TSO01}, 
and semiclassical methods \cite{SG03a}.
Another important issue is to calculate not only VER rates, but 
the physical observables related to the experiment data such as 
absorption spectrum or 2D-IR signals \cite{WH02,KDT03}. In this case,
we also need to deal with the effects of dephasing (decoherence) 
as well as VER.

The accuracy of the force field parameters is the most annoying problem.
Our preliminary calculations show that the VER rate in cyt c 
can vary by two order of magnitude when we change the bond force constant by ten percent 
(see the right of Fig.~\ref{fig:preliminary}).
This situation is rather similar to that of the reaction rate calculation,
where one must determine the activation energy accurately. 
Any inaccuracy in the activation energy causes an exponentially large 
deviation in the rate constant.
This problem will be solved through {\it ab initio} quantum dynamics (Sec.~\ref{sec:quantum}) 
or the reparametrization of the force field using experimental data or accurate {\it ab initio} calculations. 
Given sufficient accuracy in the force field, we will be in a position to discuss the relation between 
the VER and function of a protein such as electron transfer in cyt c.
As is well known, the dynamics of proteins related to function are well described 
by large amplitude (and low frequency) principal components \cite{KHG98,KHN02}. 
The connection between principal components and VER should be investigated. 
The ergodic measure \cite{ST93} will be a good device to examine this issue.
As suggested by experiments \cite{LLKH94,MK97},
collective motions in proteins can be important for the fast VER in proteins.
The collective motions near the protein surface including solvation dynamics of water 
\cite{VRPK00,HSSNK01,TT02,FFMP02,MS03} might be relevant for the VER and function.
Cytochrome c and Mb remain excellent target proteins to investigate these fundamental issues of protein dynamics and its relation to function.

\section{Acknowledgements}

We thank Dr.~Lintao Bu
for collaboration,
and Dr.~Matt Cremeens for DFT calculations.
We are grateful to the National Science Foundation (CHE-0316551) and Boston University's Center for Computational Science for generous support of our research.

\end{document}